\documentclass[aps,prd,twocolumn,showpacs]{revtex4-1}

\usepackage{bm,graphicx,textcomp,amssymb,amsmath,dcolumn,color}

\renewcommand{\Im}{\operatorname{Im}}

\newcommand{\ket}[1]{\left|#1\right>}
\newcommand{\bra}[1]{\left<#1\right|}

\newcommand{\nn}{\nonumber\\}

\newcommand{\f}[1]{\mbox{\boldmath$#1$}}
\newcommand{\fk}[1]{\mbox{\boldmath$\scriptstyle#1$}}
\newcommand{\vau}{\mbox{\boldmath$v$}}
\newcommand{\na}{\mbox{\boldmath$\nabla$}}
\newcommand{\bea}{\begin{eqnarray}}
\newcommand{\ea}{\end{eqnarray}}
\newcommand{\eea}{\end{eqnarray}}
\newcommand{\ord}{\,{\cal O}}

\begin{document}

\title{Interaction of a Bose-Einstein condensate with a gravitational wave}

\author{Ralf~Sch\"utzhold}

\affiliation{Helmholtz-Zentrum Dresden-Rossendorf, 
Bautzner Landstra{\ss}e 400, 01328 Dresden, Germany,}

\affiliation{Institut f\"ur Theoretische Physik, 
Technische Universit\"at Dresden, 01062 Dresden, Germany.}

\date{\today}

\begin{abstract}
Partly motivated by recent proposals for the detection of gravitational waves, 
we study their interaction with Bose-Einstein condensates.
For homogeneous condensates at rest, the gravitational wave does not directly 
create phonons (to lowest order), but merely affects 
existing phonons or indirectly creates phonon pairs via quantum squeezing 
-- an effect which has already been considered in the literature. 
For inhomogeneous condensate flows such as a vortex lattice, however, 
the impact of the gravitational wave can directly create phonons.
This more direct interaction can be more efficient and could perhaps help 
bringing such a detection mechanism for gravitational waves a step closer 
towards experimental realizability -- even though it is still a long way to go.
Finally, we argue that super-fluid Helium might offer some advantages in this 
respect. 
\end{abstract}


\maketitle

\section{Introduction}

A century after their prediction \cite{Einstein-1916,Einstein-1918}, 
gravitational waves have been detected at LIGO \cite{LIGO-paper,LIGO-www}, 
which was one of the major breakthroughs in modern physics. 
To fully exploit this new window into our Universe, there have been many 
proposals for alternative gravitational wave detectors, some on larger 
scale (such as LISA \cite{LISA}) and others on smaller scale. 

Pushing this idea to the extreme limit of very small systems, 
there has been a proposal \cite{Sabin} based on the creation of 
phonons in Bose-Einstein condensates by the gravitational wave.
Note that this scheme is somewhat different from an interferometric 
set-up as used in LIGO, which measures the deformation {\em during} 
the gravitational wave (e.g., with light or matter waves).
Instead, the scheme proposed in \cite{Sabin} envisions detecting the 
created phonons {\em after} the gravitational wave passed through -- 
which is more similar to resonant mass antennas such as Weber bars. 

On the one hand, the smallness of Bose-Einstein condensates made of 
atomic vapor seems to suggest that their interaction with gravitational 
waves is extremely tiny -- but, on the other hand, one might hope that 
the specific properties of Bose-Einstein condensates such as their 
coherence could help detecting this tiny interaction.
In the following, we try to adopt an unbiased point of view and address the 
general question of how Bose-Einstein condensates interact with gravitational 
waves and whether this interaction could, at least in principle, be employed 
to detect them.   

To this end, we start with a fully relativistic effective description 
of Bose-Einstein condensates in flat space-time, see also \cite{Fagnocchi}, 
and derive the non-relativistic limit in Sec.~\ref{Bose-Einstein condensate}.
Then, after briefly reviewing the well-known derivation of the phonon modes 
in flat space-time (see Sec.~\ref{Phonons}), we consider the impact of the 
gravitational wave in Sec.~\ref{Gravitational wave}.  
In Section~\ref{Scaling ansatz}, an analogy to an effective electric field 
based on an alternative scaling ansatz is introduced. 
Finally, we discuss the detectability of this effect in 
Sec.~\ref{Detectability}.  

\section{Bose-Einstein condensate}\label{Bose-Einstein condensate} 

We consider bosonic atoms with total spin zero and neglect their internal 
structure, treating them effectively as point particles. 
Since the total atom number is of course conserved, we describe them by a 
complex Klein-Fock-Gordon field $\phi$ carrying a conserved current.
In order to model the local interaction of the atoms (in the s-wave 
scattering approximation), we consider a $\lambda\phi^4$ self-interaction. 
However, as we shall see below, other suitable interactions terms yield 
the same results (in the non-relativistic limit considered here). 
Assuming that a large number of atoms is condensed into the same 
single-particle quantum state, we treat $\phi$ as a classical complex 
scalar field. 
Altogether, we start with the Klein-Fock-Gordon equation with self-interaction,  
see also \cite{Fagnocchi}
%
\bea
\label{KFG}
\left(
\frac{1}{c^2}\,\frac{\partial^2}{\partial t^2}-\na^2+
\frac{m^2c^2}{\hbar^2}
+\lambda|\phi|^2
\right)\phi=0
\,,
\ea
where $m$ denotes the mass of the atoms and $\lambda$ the their 
coupling strength. 
Since the atoms are supposed to be ultra-cold, we may consider the 
non-relativistic limit and thus separate the fast temporal oscillations 
stemming from their rest energy $mc^2$ from the remaining slow 
time-dependence via the ansatz 
\bea
\label{non-relativistic-ansatz}
\phi(t,\f{r})=\psi(t,\f{r})\exp\left\{-i\,\frac{mc^2}{\hbar}\,t\right\}
\,.
\ea
Insertion of this ansatz into~\eqref{KFG} yields 
\bea
i\hbar\dot\psi+\frac{\hbar^2}{2m}\,\na^2\psi-
\frac{\lambda\hbar^2}{2m}\,|\psi|^2\psi
=
\frac{\hbar^2}{2mc^2}\,\ddot\psi 
\,.
\ea
If we neglect the tiny relativistic correction on the right-hand side,
we recover the Gross-Pitaevskii equation 
\bea
\label{Gross-Pitaevskii}
i\hbar\dot\psi=\left(-\frac{\hbar^2}{2m}\,\na^2+g|\psi|^2\right)\psi
\,,
\ea
after identifying the coupling strength $g=\lambda\hbar^2/(2m)$.  

Instead of the $\lambda\phi^4$-interaction considered above, one might 
start with a more general ansatz for the interaction term 
$j^\mu(x) W_{\mu\nu}(x-x') j^\nu(x')$ where 
$j_\mu\propto\Im(\phi^*\partial_\mu\phi)$ is the Noether current and 
$W_{\mu\nu}(x-x')$ some interaction kernel. 
Assuming that the range of this interaction is much shorter than the 
relevant length scales of the condensate, we may approximate it by a 
local term $W_{\mu\nu}(x-x')\approx W_{\mu\nu}^0\delta^4(x-x')$, which 
is analogous to the s-wave scattering approximation. 
Then, after inserting the ansatz~\eqref{non-relativistic-ansatz}, 
we see that the rest mass density $j_0=\varrho c\propto mc$ dominates 
in the non-relativistic limit and thus we obtain the same results as 
with the $\lambda\phi^4$-coupling. 

\section{Phonons}\label{Phonons} 

Now, before considering gravitational waves, let us briefly recapitulate the 
standard derivation of the phonon wave equation from~\eqref{Gross-Pitaevskii}. 
To this end, it is convenient to employ the eikonal (WKB) ansatz 
\bea
\psi(t,\f{r})=\sqrt{\varrho(t,\f{r})}\,\exp\left\{iS(t,\f{r})/\hbar\right\}
\,,
\ea
which expresses $\psi$ in terms of density $\varrho$ and phase $S$. 
Inserting this ansatz into~\eqref{Gross-Pitaevskii}, we obtain the 
equation of continuity with the velocity $\vau=\na S/m$ 
\bea
\label{continuity}
\dot\varrho+\na\cdot(\varrho\vau)=0
\,,
\ea
and the Hamilton-Jacobi (eikonal) equation 
\bea
\label{eikonal}
\dot S+g\varrho+\frac{(\na S)^2}{2m}
=
\frac{\hbar^2}{2m}\,\frac{\na^2\sqrt{\varrho}}{\sqrt{\varrho}}
\,,
\ea
with the so-called quantum pressure term on the right-hand side.  
Neglecting this term, we obtain the Bernoulli equation for the 
condensate.  

In order to study phonons, we linearize $\varrho=\varrho_0+\delta\varrho$ 
and $S=S_0+\delta S$ these expressions around a given background solution 
$\varrho_0$ and $S_0$. 
Linearizing the equation of continuity~\eqref{continuity} yields 
\bea
\label{linear-continuity}
\left(\partial_t+\na\cdot\vau_0\right)\delta\varrho
+\na\cdot\left(\frac{\varrho_0}{m}\,\na\delta S\right)
=0
\,,
\ea
and similarly for the eikonal equation~\eqref{eikonal}
\bea
\label{linear-eikonal}
\left(\partial_t+\vau_0\cdot\na\right)\delta S+g\delta\varrho
=
\frac{\hbar^2}{4m}
\frac{\varrho_0\na^2(\delta\varrho/\sqrt{\varrho_0})-
\delta\varrho\na^2\sqrt{\varrho_0}}{\varrho_0^{3/2}}\,.
\nn
\ea
If we again neglect the quantum pressure term on the right-hand side,
we find the wave equation for sound 
\bea
\left(\partial_t+\na\cdot\vau_0\right)
\left(\partial_t+\vau_0\cdot\na\right)\delta S
=
\na\cdot\left(\frac{g\varrho_0}{m}\,\na\delta S\right)
\,,
\ea
with the convective (co-moving) derivative and the speed of sound 
$c^2_{\rm s}=g\varrho_0/m$. 

\section{Gravitational wave}\label{Gravitational wave} 

Now we are in the position to investigate how the above derivations change
in the presence of a gravitational wave. 
We employ the usual transverse trace-less (TT) gauge and consider a wave with 
a fixed (+) polarization, propagating in $z$-direction 
\bea
\label{polarization}
ds^2=c^2dt^2-\left[1-h\right]dx^2-\left[1+h\right]dy^2-dz^2
\,,
\ea
where the strain field $h(t,z)=h(t-z/c)$ describes the gravitational wave
and is very small, e.g., $h=\ord(10^{-21})$. 
The metric determinant simply reads $|g|=c^2(1-h^2)$, and we shall neglect 
all terms of second and higher order in $h$ in the following. 
As a result, the only change induced by the gravitational wave will be the 
modification of the Laplacian 
$\na^2\to[1+h]\partial_x^2+[1-h]\partial_y^2+\partial_z^2$
in~\eqref{KFG} and accordingly in the subsequent equations. 

Now, a vital point is the choice of the background solution. 
As one option, one could expand around a background solution 
$\varrho_0(t,\f{r})$ and $S_0(t,\f{r})$ in the presence of 
a gravitational wave, which may be time-dependent in general. 
In this way, one would obtain a homogeneous wave equation for sound 
whose coefficients are altered a bit due to the presence of the 
gravitational wave. 
This route has been taken in \cite{Visser} for a general metric, 
showing that the phonon propagation can be understood in terms of 
an effective acoustic metric, see also \cite{Unruh-analogue,Moncrief}. 
These results of \cite{Visser} were used in \cite{Sabin} in order to 
propose a detection mechanism for gravitational waves. 

However, the change of the condensate itself $\varrho_0(t,\f{r})$ 
and $S_0(t,\f{r})$ due to the gravitational wave is not captured 
in this homogeneous wave equation for sound.
But this change can be interpreted as a direct creation of phonons 
-- instead of merely affecting (e.g., amplifying) already existing phonons 
-- which could be a more direct signature of the gravitational wave.
Hence, we compare the two scenarios with and without the gravitational wave 
and thus linearize around a 
background solution in flat space-time (i.e., without a gravitational wave), 
e.g., the ground state $\varrho_0(\f{r})$ and $\vau_0(\f{r})$. 
Any departure from this background
solution can then be interpreted as the creation of a quasi-particle 
(e.g., phonon) and thereby a signature of the gravitational wave. 

Following this strategy, Eq.~\eqref{linear-continuity} now reads 
\bea
\label{wave-continuity}
\left(\partial_t+\na\cdot\vau_0\right)\delta\varrho
+\na\cdot\left(\frac{\varrho_0}{m}\,\na\delta S\right)
\nn
=
h\partial_y(\varrho_0 v_0^y)-h\partial_x(\varrho_0 v_0^x)
\,,
\ea
where the scalar product is taken with respect to the Minkowski metric. 
Similarly Eq.~\eqref{linear-eikonal} acquires a source term and becomes,
again after neglecting the quantum pressure term 
\bea
\label{wave-eikonal}
\left(\partial_t+\vau_0\cdot\na\right)\delta S+g\delta\varrho
=
mh\,\frac{(v_0^y)^2-(v_0^x)^2}{2}
\,.
\ea
As a result, both equations now have source terms $\propto h$ 
which enter the wave equation for sound 
\bea
\label{wave-sound}
\left(\partial_t+\na\cdot\vau_0\right)\left(\partial_t+\vau_0\cdot\na\right)
\delta S
-
\na\cdot\left(\frac{g\varrho_0}{m}\,\na\delta S\right)
\nn
=
m\left(\partial_t+\na\cdot\vau_0\right)h\,\frac{(v_0^y)^2-(v_0^x)^2}{2}
\nn
+gh\left[\partial_x(\varrho_0 v_0^x)-\partial_y(\varrho_0 v_0^y)\right]
\,.
\ea
The source terms stemming from the quantum pressure contribution are 
a bit more lengthy, but can be derived in complete analogy. 

For a homogeneous ($\varrho_0=\rm const$) condensate at rest ($\vau_0=0$), 
all the source terms vanish and thus no phonons are directly created by the 
gravitational wave.
Note that the impact on already existing phonons corresponds to terms of 
higher order $\ord(h\delta S)$ which are neglected in our first-order 
treatment. 
These higher-order terms can also induce the indirect creation of phonon pairs 
via quantum squeezing, see, e.g., \cite{Sabin}.

In a non-trivial velocity field $\vau_0(\f{r})$ such as a vortex lattice,
however, these source terms are non-vanishing and thus the gravitational
wave could directly generate phonons. 
Note that a vortex lattice is also advantageous from another point of view:
Since gravitational waves are shear waves, a medium or set-up with resistance 
or response to shear is usually considered to be favorable for gravitational 
wave detection, see, e.g., \cite{Unruh-shear,Unruh-linear}.
But a fluid has per definition no resistance to shear 
(on large length and time scales) by itself. 
However, a vortex lattice does induce a resistance to shear and thus can 
also support shear waves, which are referred to as Tkachenko waves, 
see, e.g., \cite{Tkachenko-waves}. 

\section{Scaling ansatz}\label{Scaling ansatz} 

Since the gravitational wave acts as a periodic stretching and compressing of 
the $x$ and $y$ coordinates such that the total area/volume stays constant, 
it might be illuminating to consider an appropriate ansatz for the wave 
function 
$\psi(t,x,y,z)$ which is adapted to this change. 
To this end, we exploit the fact that the condensate mainly feels 
time-dependence of $h(t,z)=h(t-z/c)$ and approximate the strain field 
$h(t,z)$ by a purely time-dependent function $h(t)$. 
We use the following scaling ansatz 
\bea
\psi(t,x,y,z)
\to
\psi(t,[1-h/2]x,[1+h/2]y,z)
\,. 
\ea
The scale factors $1\mp h/2$ are chosen such that the internal derivatives 
cancel the scale factors $1\pm h$ in front of $\partial_x^2$ and $\partial_y^2$
to linear order in $h$.
As a result, $h$ disappears from the spatial derivatives, but re-enters via 
the time-derivative 
\bea
\dot\psi(t,x,y,z)
\to
\dot\psi-\frac{\dot h}{2}
\left(x\partial_x\psi-y\partial_y\psi\right)
\,.
\ea
Accordingly, the Gross-Pitaevskii equation now reads 
\bea
i\hbar\dot\psi
=
i\hbar\,\frac{\dot h}{2}\left(x\partial_x-y\partial_y\right)\psi
-
\frac{\hbar^2}{2m}\,\na^2\psi+g|\psi|^2\psi
\,,
\ea
and thus has the same form as in the presence of an effective vector potential 
$\f{A}_{\rm eff}\propto\dot h(x\f{e}_x-y\f{e}_y)$ 
corresponding to a quadrupolar electric field 
$\f{E}_{\rm eff}\propto\ddot h(x\f{e}_x-y\f{e}_y)$.  

In terms of this scaling ansatz 
(analogous to a different choice of coordinates), 
the source terms for the equation of continuity and the eikonal equation 
(again neglecting the quantum pressure contribution) are more symmetric
\bea
&&\left(\partial_t+\nabla\cdot\vau_0\right)\delta\varrho
+\nabla\cdot\left(\frac{g\varrho_0}{m}\,\nabla\delta S\right)
=
\frac{\dot h}{2}\left(x\partial_x-y\partial_y\right)\varrho_0
\nn
&&\left(\partial_t+\vau_0\cdot\nabla\right)\delta S+g\delta\varrho
=
\frac{\dot h}{2}\left(x\partial_x-y\partial_y\right)S_0
\,.
\ea
Again we find source terms for the wave equation of sound which vanish 
for homogeneous condensates at rest.

\section{Detectability}\label{Detectability} 

Now, what remains is the important question of whether this effect is actually 
detectable. 
After all, the strength $h$ is extremely tiny, suppressed by twenty orders of 
magnitude or more. 

Let us first briefly review how this problem is solved at LIGO.
Of course, LIGO is a highly complex machine with a very clever design, 
but we shall greatly simplify our consideration by just counting how many 
orders of magnitude are gained by which main mechanisms. 
First, LIGO exploits a huge ratio of different length scales.
The arm length of the interferometer (4~km) in comparison to the wavelength 
of light (around 1~$\mu$m) gives nine orders of magnitude. 
The second source for large numbers is the ratio of time scales. 
During one half-period of the gravitational wave (in the 10~ms range),
the light bounces back and forth between the mirrors several hundred times, 
which also increases the accuracy. 
Finally, the huge number of photons within the interferometer 
(in the $\ord(10^{19})$ regime), together with our ability to detect light 
down to the single-photon limit, renders it possible to measure 
position changes of a tiny fraction of the photon wavelength, 
which accounts for the remaining orders of magnitude. 

After this brief reminder, let us discuss if one could possibly bridge this 
gap of more than twenty orders of magnitude with a Bose-Einstein condensate.
As discussed above, the lowest-order interaction Hamiltonian describing the 
coupling to a gravitational wave $h$ reads 
\bea
\label{interaction-Hamiltonian}
\hat H_{\rm int}=\hbar^2\int d^3r\,
\frac{
(\partial_x\hat\Psi^\dagger)(\partial_x\hat\Psi)- 
(\partial_y\hat\Psi^\dagger)(\partial_y\hat\Psi)
}{2m}
\,h\,. 
\ea
Since the laws of quantum mechanics imply that one can only distinguish 
orthogonal quantum states with certainty, an unambiguous detection of a 
gravitational wave is only possible if the quantum state $\ket{\psi}$ without a 
gravitational wave is orthogonal to the state $\hat U_{\rm int}\ket{\psi}$
after the interaction with the gravitational wave.
In other words, the ``no-signal'' fidelity \cite{fidelity}  
\bea
\label{fidelity}
\bra{\psi}\hat U_{\rm int}\ket{\psi}
=
\bra{\psi}
{\cal T}\exp\left\{
-\frac{i}{\hbar}\int dt\,\hat H_{\rm int}(t)\right\}
\ket{\psi}
\,,
\ea
should be zero or at least well away from 
unity (for a reasonable detection probability), see also the Appendix.
This is only possible if the smallness of $h=\ord(10^{-21})$ 
in~\eqref{interaction-Hamiltonian} is compensated by some large number(s). 

As a major advantage of a Bose-Einstein condensate, the field operator 
$\hat\Psi$, after acting on the coherent condensate state $\ket{\psi}$, 
does indeed generate a large number 
\bea
\label{c-number}
\hat\Psi(\f{r})\ket{\psi}\approx\psi_{\rm cond}(\f{r})\ket{\psi}
=
\ord(\sqrt{N})
\,,
\ea
where $\psi_{\rm cond}(\f{r})$ is the condensate wave function, 
which scales with the square root of the number of condensed atoms $N$. 
As a result, the interaction Hamiltonian~\eqref{interaction-Hamiltonian} 
scales with $\ord(Nh)$.
This enhancement mechanism can be understood via the following simple 
picture:
Neglecting the interactions between the atoms, we may approximate the 
$N$-particle wave function of the condensate by the product ansatz 
$\psi_N(\f{r}_1,\dots,\f{r}_N)=\psi_1(\f{r}_1)\dots\psi_1(\f{r}_N)$, 
where $\psi_1(\f{r})$ is the single-atom wave function, related to the 
condensate wave function via $\psi_{\rm cond}(\f{r})=\sqrt{N}\psi_1(\f{r})$. 
Now, if the fidelity (measuring the response to a gravitational wave) 
for a single atom is given by $1-\varepsilon$, where 
$\varepsilon\propto h$ is a small number, the fidelity~\eqref{fidelity}  
for the whole condensate would be 
$(1-\varepsilon)^N\approx1-N\varepsilon$.
This shows the advantage of the coherent state 
(see also \cite{Unruh-nondemolition-1,Unruh-nondemolition-2,Unruh-nondemolition-3}) 
of the condensate in 
comparison to $N$ incoherent atoms, for example, where one would have 
to add probabilities $\propto\varepsilon^2$ instead of amplitudes 
$\propto\varepsilon$, which gives the usual $N\varepsilon^2$ versus 
$N\varepsilon$ scaling. 

Note that the scaling $\ord(Nh)$ is due to the fact that the gravitational 
wave interacts with the whole condensate 
(as considered in the previous sections), see also the Appendix. 
If we have a homogeneous condensate at rest $\na\psi_{\rm cond}=0$ plus a 
few phonons, the gravitational wave would only act on these phonons and 
thus the scaling would be reduced to $\ord(nh)$ where $n$ is the number 
of phonons, which is much smaller $n\ll N$.
This can be understood by inserting the usual mean-field ansatz 
\bea
\label{mean-field}
\hat\Psi(\f{r})\approx\psi_{\rm cond}(\f{r})+\hat\chi(\f{r})
\,,
\ea
%
%
where $\hat\chi(\f{r})$ are the Bogoliubov-de Gennes (or phonon) modes. 
Their action on $\ket{\psi}$ scales with $\sqrt{n}$ instead of $\sqrt{N}$,
i.e., $\hat\chi\ket{\psi}=\ord(\sqrt{n})$. 

In general, for a condensate with $\na\psi_{\rm cond}\neq0$, 
the bilinear structure of the interaction 
Hamiltonian~\eqref{interaction-Hamiltonian} 
entails the following hierarchy:
To leading order $\ord(hN)$, both field operators act on $\ket{\psi}$ 
according to~\eqref{c-number}, i.e., as c-numbers, see also the 
Appendix.
The next order with an intermediate scaling $\ord(h\sqrt{nN})$ 
is caused by mixed terms where one field operator acts as a 
c-number according to~\eqref{c-number} while to other one 
creates (or annihilates) a Bogoliubov-de Gennes phonon, cf.~Eq.~\eqref{mean-field}. 
Still, these contributions are caused by the response of the whole 
condensate and vanish for homogeneous condensates at rest 
$\na\psi_{\rm cond}=0$.
Finally, if both field operators act on Bogoliubov-de Gennes 
(phonon) modes, we get the scaling $\ord(hn)$ discussed above. 

Unfortunately, Bose-Einstein condensates of ultra-cold atoms do typically 
not contain enough atoms to compensate twenty orders of magnitude.
Since the characteristic length and time scales of such condensates are 
usually in the $\mu$m and ms regime, it is also not easy to generate further 
large numbers by ratios of length or time scales. 
One option (also discussed in \cite{Sabin}) could be based on resonance 
effects, 
which would require a sufficiently long life-time of the condensate with 
a high enough Q factor for the relevant modes as well as a gravitational 
wave with precisely the right frequency (which must also be stable over 
that time), see also the Appendix. 
In view of these obstacles, overcoming the twenty orders of magnitude 
would require tremendous experimental progress and new ideas. 

In addition, achieving a sufficiently small fidelity~\eqref{fidelity}  
is not the end of the story.
This just implies that the laws of quantum mechanics do not forbid the 
detection of gravitational waves via this mechanism.
To actually measure the difference between the states $\ket{\psi}$ and 
$\hat U_{\rm int}\ket{\psi}$, e.g., to detect single phonons is a highly 
non-trivial task (see, e.g., \cite{Detection}).
This shows another advantage of LIGO, because our experimental capabilities 
to detect light (in the optical or near-optical regime) down to the 
single-photon level is well developed. 

Even with being able to detect single phonons, there is still the task of 
distinguishing the phonons created by a gravitational wave from other noise 
effects.  
In this regard, it is important to remember a crucial difference between 
LIGO and the scheme discussed here: 
While LIGO measures the position changes of the mirrors while the 
gravitational wave passes through, one would detect phonons in the 
condensate {\em after} the interaction with the gravitational wave.
From this point of view, the condensate is more analogous to resonant 
mass antennas such as Weber bars (as mentioned in the Introduction). 
As a result, the matching to gravitational wave form templates as used 
in LIGO cannot be applied in the same way here, which makes it necessary 
to employ other mechanisms to filter out the noise. 

To end this Section with some speculations, one might consider using 
super-fluid Helium instead of ultra-cold atomic vapor, see also 
\cite{Helium}. 
As a drawback, super-fluid Helium is a strongly interacting system which 
is harder to model theoretically and only a small fraction of the atoms 
(a bit below 10~\%) are actually condensed. 
As an advantage, however, the number of atoms and thus also the phase space 
for length and time scales can be much larger. 
Of course, the issues related to detecting single or a small number of 
phonons or Tkachenko quanta and filtering out noise are analogous. 

\section{Conclusions \& Outlook} 

Starting with a fully relativistic effective description~\eqref{KFG} of 
a Bose-Einstein condensate, we study its interaction with a gravitational 
wave~\eqref{polarization}. 
We find that a vital point is the choice of the background solution 
around which the phonon modes are linearized. 
In order to calculate the phonons (or other quasi-particle excitations) 
directly created by the gravitational wave, we choose a background solution 
$\f{v}_0$ and $\varrho_0$ in flat space-time, i.e., without a gravitational wave.
Then we obtain an inhomogeneous wave equation~\eqref{wave-sound} 
\bea
\label{inhomogeneous}
\Box_{\fk{v}_0,\varrho_0}\delta S={\mathfrak D}_{\fk{v}_0,\varrho_0}h
\,,
\ea
for the phonons $\delta S$ where the gravitational wave $h$ generates a source 
term -- unless we have a homogeneous condensate at rest. 
In contrast, if we choose a background solution $\f{v}_g$ and $\varrho_g$ in the 
presence of a gravitational wave
(or, equivalently, if we have a homogeneous condensate at rest), 
we obtain a homogeneous wave equation
\bea
\label{homogeneous}
\Box_{\fk{v}_g,\varrho_g}^h\delta S=0
\,,
\ea
where the gravitational wave only modifies the coefficients a little bit,
see, e.g., \cite{Sabin,Visser}.
This modification is of higher order $\ord(h\delta S)$ than the effect 
in~\eqref{inhomogeneous} which is linear in both, $h$ and $\delta S$. 

In other words, we obtain a hierarchy of effects
(see also \cite{Raetzel} for a different scenario).  
To leading order, the gravitational wave can directly create phonons 
(in \cite{Raetzel} referred to as direct driving)
from the inhomogeneous condensate flow according to the
inhomogeneous wave equation~\eqref{inhomogeneous}. 
This effect scales (at most) with the total number $N$ of atoms 
in the condensate $\ord(Nh)$, see also the Appendix. 
To sub-leading order, the gravitational wave can affect 
(e.g., amplify or de-amplify) already existing phononic excitations 
(in \cite{Raetzel} referred to as parametric driving) 
according to the inhomogeneous wave equation~\eqref{homogeneous}. 
These phonons could be in a Fock $\ket{n}$, 
a coherent $\ket{\alpha}$, or a squeezed state $\ket{\xi}$,
for example. 
For a coherent state $\ket{\alpha}$, one can re-interpret these 
phononic excitations as small variations  
$\psi_{\rm cond}\to\psi_{\rm cond}+\delta\psi_{\rm cond}$
of the condensate wave-function, i.e., small density and phase 
fluctuations. 
Effectively, this corresponds to inhomogeneities of the condensate 
which can be regarded as small source terms in Eq.~\eqref{wave-sound}.
Since the coherent states $\ket{\alpha}$ form a complete basis,
other states such as Fock $\ket{n}$ or squeezed states $\ket{\xi}$
can be interpreted as superpositions of states with different 
variations $\delta\psi_{\rm cond}$.
Consistent with this picture, the sub-leading effect scales with 
the number $n$ of phonons in the condensate $\ord(nh)$ which is 
typically much smaller \cite{footnote}.
Finally, even if no phonons are present initially, the 
gravitational wave can create phonons out of the quantum 
ground state fluctuations (the quantum depletion of the condensate).
For this spontaneous effect, the scaling with $\ord(n)$ has to be 
replaced by order unity, i.e., this effect scales with $\ord(h)$.
Altogether, we have the hierarchy 
(including the mixed terms, cf.~Sec.~\ref{Detectability}) 
\bea
\ord(hN)\gg\ord(h\sqrt{nN})\gg\ord(hn)\gg\ord(h)\,.
\ea

Nevertheless, since $hN$ is still a tiny number in typical Bose-Einstein 
condensates made of atomic vapor, further large numbers 
(such as ratios in length and time scales, as in LIGO) 
would be required to reach the regime necessary for gravitational 
wave detection. 
In view of the characteristic length and time scales of typical 
Bose-Einstein condensates made of atomic vapor, this seems to be 
an extremely challenging task (see also the Appendix). 
As a speculation, this task might be a bit less challenging for 
super-fluid Helium, where the number $N$ and also the spatial 
dimensions can be much larger. 
In addition, super-fluid Helium is the real ground state of the system 
instead of the meta-stable state of Bose-Einstein condensates made of 
atomic vapor (which entails problems with three-body losses etc.).

It is also important to note that we did not include a potential $V$
(such as the trapping potential) in our considerations. 
Its interaction Hamiltonian reads 
\bea
\label{trap-potential}
\hat H_{\rm int}^V
=
\int d^3r\,\hat\Psi^\dagger\hat\Psi\,\frac{\partial V}{\partial h}\,h
\,.
\ea
Ideally, one should also start with a fully relativistic description  
of the potential (e.g., generated by laser beams) and then derive 
its change $\partial V/\partial h$ due to the gravitational wave. 
However, all the arguments above would still apply (at least qualitatively)
since the above Hamiltonian would only act as an additional source for 
phonons (or other excitations), it would require tremendous fine-tuning 
to have the phonons created by $\hat H_{\rm int}^V$ cancel the other 
phonons created in the condensate.  

\acknowledgments 

The author acknowledges fruitful discussions during the Analogue-Gravity 
Workshop 18-20~June 2018 at the Technion (Israel) as well as the 
The Relativistic Quantum Information -- North Conference 
24-27 September 2018 at the University of Vienna (Austria). 

\appendix
\section*{Appendix}

In contrast to the rather general order of magnitude estimates above,
let us try to obtain an explicit bound for the sensitivity of the 
considered detection scheme. 
To this end, we Taylor expand the ``no-signal'' fidelity~\eqref{fidelity}  
\bea
\label{fidelity-Taylor}
\bra{\psi}\hat U_{\rm int}\ket{\psi}
=
1-\frac{i}{\hbar}\int dt\,\bra{\psi}\hat H_{\rm int}(t)\ket{\psi}
+\ord(h^2)
\,.
\ea
We see that the first-order correction in $h$ is purely imaginary, 
i.e., corresponds to a pure phase shift, which can be interpreted 
as a global phase fluctuation $\delta S$.
For an isolated quantum system, such a pure phase shift would not be 
measurable, but in comparison with another quantum system the relative 
phase shift $\delta S$ could be measured -- at least in principle. 
As one possibility, one could imagine a quantum superposition state 
where either all $N$ atoms are in the condensate which interacts with 
the gravitational wave or all $N$ atoms are in another condensate 
which does not interact with the gravitational wave (e.g., due to 
a different geometry). 
Such highly non-classical (entangled) states $(\ket{N,0}+\ket{0,N})/\sqrt{2}$ 
are often referred to as NOON states, see also \cite{noon}. 

Inserting the interaction Hamiltonian~\eqref{interaction-Hamiltonian},  
the leading-order phase shift from~\eqref{fidelity-Taylor} reads 
\bea
\label{phase-shift}
\varphi 
=
-\frac1\hbar\int dt\,\bra{\psi}\hat H_{\rm int}(t)\ket{\psi}
=
-\frac1\hbar\int dt\,h(t)\times
\nn
\hbar^2\int d^3r\,
\bra{\psi}
\frac{
(\partial_x\hat\Psi^\dagger)(\partial_x\hat\Psi)- 
(\partial_y\hat\Psi^\dagger)(\partial_y\hat\Psi)
}{2m}
\ket{\psi}
\,,
\ea
where we have neglected the spatial dependence of $h$, 
as explained in Sec.~\ref{Scaling ansatz}.  
For an oscillatory time-dependence of $h(t)$ and a 
stationary condensate state $\ket{\psi}$, 
the time integral would vanish. 
Thus, in order to obtain a leading-order phase 
shift~\eqref{fidelity-Taylor}, 
one should have a non-oscillatory (e.g., peaked) 
time-dependence $h(t)$ or a non-stationary condensate 
state $\ket{\psi}$.
Considering the latter possibility, the time integral 
would grow linearly with measurement time $T$ if the 
condensate motion is in resonance with the 
(sinusoidal) gravitational wave.  
Since the characteristic time scales of typical condensates 
are in the ms regime or longer, this would correspond 
to gravitational wave frequencies in the kHz range or below
(cf.~\cite{LIGO-paper}). 

The remaining spatial integral in the second line of 
Eq.~\eqref{phase-shift} can be bounded from above by the total 
kinetic energy $E_{\rm kin}$ of the condensate, because 
$E_{\rm kin}$ is given by the same integral over the 
sum of the two non-negative terms 
$\bra{\psi}(\partial_x\hat\Psi^\dagger)(\partial_x\hat\Psi)\ket{\psi}$
and 
$\bra{\psi}(\partial_y\hat\Psi^\dagger)(\partial_y\hat\Psi)\ket{\psi}$
plus the same expression in $z$-direction. 
Finally, since the interaction energy of the condensate is positive 
(for repulsive interactions), the kinetic energy $E_{\rm kin}$ 
is always smaller than the total energy $E$ of the condensate. 
Altogether, the leading-order phase shift~\eqref{phase-shift}
can be strictly bounded from above by 
\bea
\label{bound}
|\varphi|
\leq
\frac{T h_{\rm max} E}{\hbar}  
\,.
\ea
Since the total energy $E$ scales with the number $N$ of atoms, we recover 
the scaling $\ord(hN)$ discussed above. 

In order to get a feeling for the involved orders of magnitude, 
let us consider a gravitational wave with a rather large amplitude 
$h_{\rm max}=10^{-21}$ (cf.~\cite{LIGO-paper}) and precisely the 
correct resonance frequency, which is also assumed to be stable over a 
comparably long integration/measurement time of $T=2000~\rm s$ 
(cf.~\cite{Sabin}). 
In order to obtain a measurable phase shift $\varphi$ 
not too far below unity, the total energy $E$ of the condensate should 
exceed 100~eV, which is a huge energy. 
Assuming an atom number of $N=10^6$ (cf.~\cite{Sabin}),
this corresponds to a temperature scale of order Kelvin, 
which is too large in comparison to typical condensation 
temperatures below $\mu$K. 
(Note that the temperature scales for super-fluid Helium 
are much higher, which shows another potential advantage of this 
system.)

We may also turn this argument around and consider characteristic 
time and energy scales of typical condensates. 
For $N=10^6$ atoms and typical velocities of order mm/s 
(corresponding to a temperature scale in the 100~nK regime), 
we obtain a kinetic energy $E_{\rm kin}$ of several $\mu$eV.
Assuming an integration/measurement time of $T=0.1~\rm s$ 
(cf.~\cite{LIGO-paper}), this would allow us to detect, 
at least in principle,  
gravitational waves with $h_{\rm max}\geq\ord(10^{-9})$. 
Note, however, that this detection scheme is based on the highly 
non-classical NOON states mentioned above. 
This sensitivity $h_{\rm max}\geq\ord(10^{-9})$ stems mostly 
from the large number $N=10^6$ of atoms, the remaining orders  
of magnitude come from the integration/measurement time 
(similar to the discussion in Sec.~\ref{Detectability}). 

Since the density $\hat\varrho=\hat\Psi^\dagger\hat\Psi$ is 
non-negative, an analogous upper bound can be derived for the  
interaction Hamiltonian~\eqref{trap-potential} stemming from 
the trap potential
\bea
\label{bound-trap} 
|\varphi|
\leq
\frac{T h_{\rm max} N}{\hbar}\,  
\left|\frac{\partial V}{\partial h}\right|_{\rm max}
\,,
\ea
where we directly see the scaling $\ord(hN)$. 
The arguments and typical energy and time scales are very 
similar to those discussed after~\eqref{bound}, but one might 
hope to obtain an additional amplification factor by using 
steep potentials $V$ such that $\partial V/\partial h$
becomes large -- which is analogous to the ratio of length 
scales discussed in Sec.~\ref{Detectability}. 



\begin{thebibliography}{1}
\bibitem{fidelity}
Note that there is a potential source for confusion here:
%
Usually, a fidelity close to unity is desirable 
(e.g., for quantum control or quantum algorithms).
%
In contrast, for the detection scheme considered here, 
a fidelity away from unity (ideally close to zero) is desirable. 
%
In order to avoid this potential source for confusion,
we refer to~\eqref{fidelity} as ``no-signal'' fidelity.
\bibitem{footnote}
Note that by considering large squeezing parameters such as $\xi=10$ 
and a number $N=10^6$ of atoms (as in \cite{Sabin}), one would formally 
obtain extremely large phonon numbers $\langle\hat n\rangle=\sinh^2\xi>10^8$ 
which even exceed $N$. 
%
However, in this regime, the applicability of the linearized phonon description 
should be called into question. 
%
In addition, one should avoid double counting, which could feign a scaling with 
$\ord(nNh)$.
\end{thebibliography}
\end{document}